\documentclass[12pt]{article}
\usepackage{amsmath,amssymb,epsfig}

\usepackage{color}
\def\unit{{\relax{\rm 1\kern-.26em I}}}

\newcommand{\beq}{\begin{equation}}
\newcommand{\eeq}{\end{equation}}
\newcommand{\beqa}{\begin{eqnarray}}
\newcommand{\eeqa}{\end{eqnarray}}
\newcommand{\simg}{\gtrsim}
\newcommand{\siml}{\lesssim}

\newcommand{\CO}{{\cal O}}

\newcommand{\al}[1]{\begin{align}#1\end{align}}

\newcommand{\red}{\textcolor{black}}
\newcommand{\blue}{\textcolor{black}}
\newcommand{\add}{\textcolor{black}}

\newcommand{\nn}{\nonumber\\}
\newcommand{\GeV}{\ensuremath{\,\text{GeV} }}

\newcommand{\TeV}{\ensuremath{\,\text{TeV} }}

\makeatletter

\renewcommand\section{\@startsection {section}{1}{\z@}%
                                   {-3.5ex \@plus -1ex \@minus -.2ex}%
                                   {2.3ex \@plus.2ex}%
                                   {\normalfont\large\bfseries}}

\renewcommand\subsection{\@startsection{subsection}{2}{\z@}%
                                     {-3.25ex\@plus -1ex \@minus -.2ex}%
                                     {1.5ex \@plus .2ex}%
                                     {\normalfont\normalsize\bfseries}}

\newcount\hour \newcount\minute
\hour=\time \divide \hour by 60
\minute=\time
\def\now{%
\ifnum \hour<13
  \ifnum \hour=0 \advance \hour by 12 \number\hour:\else \number\hour:\fi%
     \ifnum \minute<10 0\fi%
     \number\minute%
\ A.M.%
\else \advance \hour by -12 \number\hour:%
  \ifnum \minute<10 0\fi%
  \number\minute%
  \ P.M.%
\fi%
}

\makeatother

\begin{document}

\baselineskip=18pt  
\numberwithin{equation}{section}  
\allowdisplaybreaks  



%
%


\thispagestyle{empty}

\vspace*{-2cm}
\begin{flushright}
\end{flushright}

\begin{flushright}
KUNS-2465\\
DESY 13-170 \\
\end{flushright}

\begin{center}

\vspace{1.4cm}

\vspace{0.5cm}
{\bf \Large More on cosmological constraints on spontaneous R-symmetry breaking models}
\vspace*{0.2cm}

\vspace{0.5cm}

{\bf
Yuta Hamada$^{1}$, Kohei Kamada$^{2,\blue{3}}$, Tatsuo
Kobayashi$^{1}$ and Yutaka Ookouchi$^{4}$}
\vspace*{0.5cm}

\vspace*{0.5cm}

$^{1}${\it Department of Physics, Kyoto University, Kyoto 606-8502, Japan}\\

\vspace{0.1cm}

$^2$ {\it Deutsches Elektronen-Synchrotron DESY,
Notkestrasse 85, D-22607 Hamburg, Germany }\\

\vspace{0.1cm}\blue{
$^{3}${\it Institut de Th\'eorie des Ph\'enom\`enes Physiques,
\'Ecole Polytechnique F\'ed\'erale de Lausanne,
CH-1015 Lausanne, Switzerland}}\\
\vspace{0.1cm}

$^{4}${\it Faculty of Arts and Science, Kyushu
 University, Fukuoka 819-0395, Japan  }\\

\vspace*{0.5cm}

\end{center}

\vspace{1cm} \centerline{\bf Abstract} \vspace*{0.5cm}
\red{
We study the spontaneous R-symmetry breaking model and investigate the
cosmological constraints on this model due to the pseudo
Nambu-Goldstone boson, R-axion. We consider the R-axion which has
relatively heavy mass in order to complement our previous work. In this
regime, model parameters, R-axions mass and R-symmetry breaking
  scale,
are constrained by Big Bang Nucleosynthesis and overproduction of the gravitino produced from R-axion decay and thermal plasma.}
\blue{We find that the allowed parameter space is very small for 
high reheating temperature.
For low reheating temperature, 
the $U(1)_R$ breaking scale $f_a$ is constrained as
$f_a<10^{12-14}\GeV$ regardless of the value of R-axion mass.}

\newpage
\setcounter{page}{1} 



\section{Introduction}

Supersymmetry (SUSY) is one of the most promising candidates of the physics beyond the standard model (SM) because 
it can relax the naturalness problem and suggests the gauge coupling unification.
{}It is also widely believed to be one of the key ingredients for constructing a consistent string theory encompassing the SM. Since SUSY has not been observed in experiments yet, it has to be broken somewhere between the weak scale and the Planck scale. Recent discovery of a Higgs boson at the LHC \cite{LHChiggs} may suggest that 
the stop mass is around ${\cal O}(10)$ TeV without introducing an artificial new mechanism \cite{SUSYHiggs}, in particular, in models of gauge mediation, which is the main topic in this paper.
Although it is not at the right scale for solving the naturalness
problem,\blue{\footnote{\blue{In a certain scenario, a heavy stop mass such
    as 
several TeV is still natural \cite{Choi:2005hd}.}}} 
other good points of SUSY encourage us to study it further. 
Therefore,  we focus on relatively high-scale SUSY-breaking scenarios in this paper.

If high-scale SUSY is realized in nature, it would be interesting to
seek for a connection between the SUSY breaking scale and cosmological observations, which are quite useful tools to probe the high-scale physics beyond the TeV scale.
Among many other scenarios of high-scale SUSY breaking, gauge mediation models with spontaneously broken R-symmetry, which is a specific symmetry in supersymmetric theories, is one of \red{the models that has recently experienced striking progress} on model building \cite{Shih1}, (see for reviews \cite{SUSYrev1,SUSYrev2,SUSYrev3}). 
As is emphasized in
Refs.~\cite{Hamada:2012fr,Eto:2012ij,Rthermal,Hiramatsu}, R-symmetry
opens up interesting windows into the connection \blue{between 
SUSY breaking and cosmological aspects}. In particular, cosmology with the Nambu-Goldstone boson, called R-axion, which is generated and acquires a mass term in coupling to the gravity theory because the constant term in superpotential breaks R-symmetry explicitly 
is an interesting working place;
R-axions are produced at some time in the cosmic history and their decays may affect the standard cosmological scenario, 
which, in turn, constrains the model parameters \cite{Hamada:2012fr}. 

Depending on its mass scale, various decay modes are allowed. 
In our previous study \cite{Hamada:2012fr}, we focused on relatively light and long-lived R-axions since such parameter regions are favored in the context of ``low-scale gauge mediation'' \cite{Goh:2008xz}, where various cosmological constraints including the Big Bang Nucleosynthesis (BBN), the Cosmic Microwave Background (CMB), cosmic $\gamma$-ray and the re-ionization can be imposed in the late epoch of the expanding universe. 
In addition to these constraints, we here point out that the abundance of heavier R-axions with shorter lifetime, which can explain the 125 GeV Higgs more easily, can be constrained in a wide range of parameter space by two cosmological constraints: 
One is coming from hadronic decays of R-axions. When the mass scale
of the R-axion is much larger than ${\cal O}(1)$ GeV, the R-axion can efficiently decay into various hadrons. 
For sufficiently heavy R-axions, they immediately turn to hadronic jets, which affects successful BBN. 
This phenomenon can constrain the parameter space of R-axions.
\red{The other constraint comes from gravitino production. In this
short note, we mainly focus on the regime in which gauge mediation is
the dominant contribution to the mediation of SUSY breaking. In this
case, the gravitino is a stable particle and can be over-produced via R-axion decay. In fact, as we will show below, overproduction of such gravitinos yields a condition which is complementary to the one for thermal production of gravitino. }

%
%
%

In this paper, we investigate cosmological effects of heavy R-axion
whose mass scale is \red{not covered in} the previous work
\cite{Hamada:2012fr}. Especially, we focus \blue{the mass scale
  heavier than $3\GeV$ } to avoid subtlety of non-relativistic decay
into pions which requires careful treatment because it does not
necessarily destroy the light elements constructed by BBN. Also, we
assume that all superparticles (except gravitino) are heavier than the
R-axion 
because R-axion decay into superparticles requires highly model dependent argument. 
If the decay channel into superparticles of R-axions opens, the constraint would become more severe.

The organization of this paper is as follows: 
In section \ref{sec2}, we briefly review the R-symmetry breaking model.  
Then, we examine the lifetime of the R-axion and the branching ratio of hadronic decays. In section \ref{sec4}, we firstly review the mechanisms of the R-axion production and its abundance. Then we impose constraints on R-axion abundance from the BBN and gravitino overproduction. As we will show that the R-axion mass and R-symmetry breaking scale are severely constrained. Section \ref{sec6} is devoted to summary and discussion.

\section{Hadronic decay of R-axion \label{sec2}}

In this section, we briefly review the simple model with spontaneously broken R-symmetry studied in Ref.~\cite{Hamada:2012fr}. 
Let us focus on the R-charged light SUSY-breaking field, $X$, and consider a low-energy scale where
all other fields including messenger fields are integrated out. The effective superpotential is, then, assumed to be
\begin{equation}
W_{\rm eff}=\Lambda_{\rm eff}^2 X+W_0. \label{superpotential}
\end{equation}
Here the constant term $W_0$ is required to realize vanishing cosmological constant. 
Note that this class of models is 
common in various $F$-term supersymmetry breaking models \blue{\cite{Shih1}}. 
We assume non-canonical 
K\"ahler potential yielding the following potential,
\begin{align}
V(X)&=\frac{\lambda}{4}\left(|X|^2-f_a^2\right)^2-2\frac{W_0 \Lambda_{\rm eff}^2}{M_{\rm pl}^2} X+{\rm h.c.}+\cdots,   \label{potax}
\end{align}
which realizes a potential minimum with spontaneously broken R-symmetry.
Here $M_{\rm pl}$ is the reduced Planck mass.
The first term comes from our assumption about K\"ahler potential. 
$f_a$ is turned out to be the ``axion decay constant'' and we assume $f_a \ll M_{\rm pl}$.
The second term appears from the Planck-suppressed interaction in
supergravity theory.
It breaks R-symmetry explicitly and generates the mass term for the R-axion, the phase component of the $X$ field.  
In coupling to the gravity, vanishing cosmological constant requires
\al{
\Lambda_{\rm eff}^4=3 W_0^2/M_{\rm pl}^2. 
}
Though we have introduced an ad-hoc K\"ahler potential for the SUSY-braking vacuum with spontaneously broken R-symmetry,
we believe that it is a simple toy model which reveals various aspects
in spontaneous R-symmetry breaking models \footnote{\add{As we will see later, in order to put cosmological constraints, we need to know the interactions and the abundance of R-axions.
In spontaneous R-symmetry breaking models, these are characterized by R-axion mass $m_a$ and decay constant $f_a$ for any K\"ahler potential.
Therefore, taking $m_a$ and $f_a$ as free parameters, our discussion of ad-hoc K\"ahler potential is applicable to many spontaneous R-symmetry breaking models where the R-symmetry breaking is realized by general non-canonical K\"ahler potential.
}
}\blue{\cite{Shih1}}. Dividing $X$ into VEV and fluctuation, 
\al{
X=\frac{s+\sqrt{2} f_a}{\sqrt{2}} \exp (ia/\sqrt{2}f_a), 
}
we have the potential for $s$ and $a$ as
\al{\label{our potential}
V(s,a)=\frac{\lambda}{2}f_a^2 s^2-4\frac{W_0 \Lambda_{\rm eff}^2}{M_{\rm pl}^2}f_a \cos \left({\frac{a}{\sqrt{2}f_a}}\right)+\frac{1}{2\sqrt{2}}m_a^2 f_a s \cos \left({\frac{a}{\sqrt{2}f_a}}\right).
}
Hereafter we call $s$ R-saxion and $a$ R-axion. 
From the potential, we can read off the R-axion mass $m_a$ and R-saxion mass $m_s$ as
\al{
m_a^2=\frac{2W_0 \Lambda_{\rm eff}^2}{f_a M_{\rm pl}^2},\quad 
\label{saxion1}
m_s^2=\lambda f_a^2.
}
We expect that R-symmetry breaking potential  Eq.\eqref{potax} would be related to SUSY breaking and
\al{
V(0)\simeq \lambda f_a^4\simeq\langle F_X\rangle^2 
=\Lambda_{\rm eff}^4.
}
With this assumption, we find the relation of $m_a$ and $m_s$ as 
\al{\label{saxion2}
m_s^2\simeq\frac{M_{\rm pl}}{f_a}m_a^2.
}
Meanwhile the gravitino mass $m_{3/2}$ is written by
\al{
m_{3/2}^2\simeq\frac{f_a}{2\sqrt{3}  M_{\text{pl}}}m_a^2. \label{gravmass}
} 
Since the gravitino is lighter than R-axion while R-saxion is heavier than R-axion, 
R-axion can decay into gravitinos but not R-saxions. 

Now we evaluate the  interactions of R-axions, which is necessary to investigate 
the production and decay rate of R-axions. 
First, let us consider the interaction with the gauge fields. R-axions
couple with the SM gauge fields through anomaly couplings,
\al{
& \blue{\frac{C_{i}g^2_{i}}{32 \pi^2 f_a}aF^{G_i}_{\mu \nu} \tilde F^{G_i\mu \nu},
\quad {\rm with} \quad C_{i}=\mathrm{Tr}~ U(1)_R G_{i}^2,
}}
\blue{where $G_i$ for $i=3, 2$ and 1 represent the SM gauge groups
$SU(3)_C$, $SU(2)_L$ and $U(1)_Y$, respectively 
and $g_{i}$s are the corresponding gauge couplings. }
{}From these couplings, we obtain the decay rates of R-axion to each pair of gauge bosons as
\al{\label{decay width}
&\Gamma (a\rightarrow 2g)=\frac{C_{3}^2}{\red{2}\pi}\left(\frac{g_3}{4\pi}\right)^4\left(\frac{m_a}{f_a}\right)^2 m_a,\nn
&\Gamma (a\rightarrow 2\gamma)=\frac{(C_2 \sin^2\theta_w g_2^2+C_Y \cos^2 \theta_w g_Y^2)^2}{16\pi(4\pi)^4}\left(\frac{m_a}{f_a}\right)^2 m_a,\nn
&\Gamma (a\rightarrow 2Z)=\frac{1}{16\pi (4\pi)^4}(C_2 \cos^2\theta_w g_2^2+C_Y \sin^2 \theta_w g_Y^2)^2\left(\frac{m_a}{f_a}\right)^2 m_a\left(1-\frac{4m_Z^2}{m_a^2}\right)^{3/2},\nn
&\Gamma (a\rightarrow 2W)=\frac{C_{2}^2}{8\pi}\left(\frac{g_{2}}{4\pi}\right)^4\left(\frac{m_a}{f_a}\right)^2 m_a\left(1-\frac{4m_W^2}{m_a^2}\right)^{3/2},\nn
&\Gamma(a\rightarrow \gamma Z)=\frac{\cos^2\theta_w \sin^2 \theta_w}{8\pi (4\pi)^4}(C_2  g_2^2-C_Y g_Y^2)^2 \left(\frac{m_a}{f_a}\right)^2 m_a \left(1-\frac{m_Z^2}{m_a^2}\right)^{3},
}
where $\theta_w$ is the Weinberg angle and $m_Z$ and $m_W$ are the Z-boson and W-boson masses, respectively.
Since the anomaly coefficients are model-dependent parameters of the order of the unity, we take \blue{$C_{i}=1$} for all $G_i$ in the following. 
Taking other values of the order of the unity does not change our results significantly.

Secondly, the R-axion can also couple with the SM fermions through the
mixing between the R-axion and the Higgs bosons \cite{Goh:2008xz}. Couplings with up type quarks, down type quarks, charged leptons and neutrinos are expressed as the effective interactions, \red{$\lambda_f a  {\bar f}\gamma^5 f,$} with the coupling constants
\begin{eqnarray}
\lambda_u &=& i\frac{m_u}{f_a}\kappa \cos^2 \beta, \nonumber \\
\lambda_d &=& i\frac{m_d}{f_a} \kappa \sin^2 \beta,  \nonumber \\
\lambda_\ell &=& i\frac{m_\ell}{f_a}\kappa \sin^2 \beta, \nonumber \\
\lambda_\nu &=& i\frac{m_\nu}{ f_a}\kappa \cos^2 \beta,
\end{eqnarray}
where $\kappa=v/(\sqrt{2}f_a)$ with $v=246\GeV$ and $m_f$ denotes the mass of each fermion $f$. 
$\tan \beta$ is the ratio of the vacuum expectation values of the up-type Higgs boson 
$H_u$ and the down-type Higgs boson $H_d$.
From these couplings, R-axions can decay into a pair of  fermions with the decay rates, 
\al{
\Gamma(a \rightarrow f{\bar f}) = \dfrac{\left|\lambda_f\right|^2}{8 \pi} m_a 
\left( 1 - 4m_f^2/m_a^2 \right)^{1/2}\times 
 \left\{
\begin{array}{ll}
3 &\text{for}\quad f=u, d\\
1 & \text{for} \quad f=l, \nu
\end{array}\right.
.}
Note that $\tan\beta$ as well as other parameters such as the stop mass determines the Higgs mass. 
For the 125 GeV Higgs, $\tan \beta \gtrsim 10$ is favored for the stop mass with a few TeV \cite{SUSYHiggs}. 

Finally, 
R-axions can decay to a pair of gravitinos through supergravity effect, 
\begin{equation}
\frac{W^*}{M_{\rm pl}^2} \psi_\mu \sigma^{\mu\nu} \psi_\nu +{\rm h.c.} \ni -i\frac{\Lambda_{\rm eff}^2 a}{\sqrt{2}M_{\rm pl}^2}\psi_\mu\sigma^{\mu\nu} \psi_\nu+{\rm h.c.}
\end{equation}
Decay rate is given by
\al{\label{gravitino_decay}
\Gamma (a\rightarrow 2\psi)&\simeq 
\dfrac{1}{\red{2}\sqrt{3} \pi }\frac{m_a^3}{ M_{\rm{pl}}f_a}
,
}
for $m_a\gg m_{3/2}$.
Here we have used the relation $\Lambda_{\rm eff}^2 = 3^{1/4} \sqrt{f_a M_{\rm pl}/2} \ m_a$ 
and taken into account that for the light gravitino, the decay into the spin 1/2 goldstino component $\psi$ with 
$\psi_\mu \sim i \sqrt{2/3} \ \partial_\mu \psi/m_{3/2}$ dominates
over the decay into the spin 3/2 component \blue{\cite{Moroi:1995fs}}. 

\begin{figure}[h]
\begin{center}
\includegraphics[width=.5\textwidth]{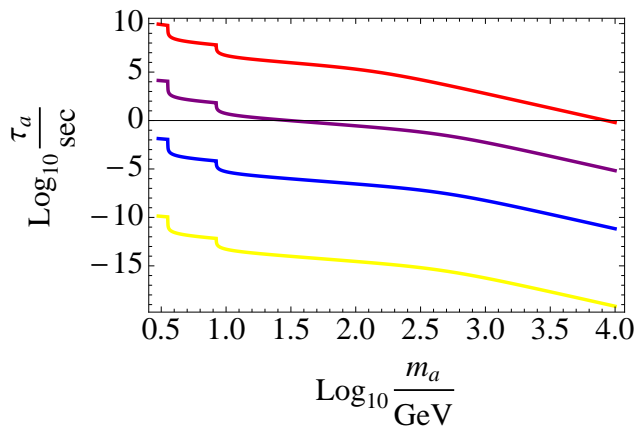}
\caption{R-axion lifetime $\tau_a$ with various values of $f_a$. Yellow, blue, purple and red lines correspond to $f_a=10^6\GeV, 10^{10}\GeV, 10^{13}\GeV,$ and $10^{16}\GeV$, respectively. Here we use $\tan\beta=30$. }\label{lifetime}
\end{center}
\end{figure}

We plot the lifetime $\tau_a$ of the R-axion  for $\tan \beta =30$ in Fig.\ref{lifetime}. 
For $3 {\rm GeV} \lesssim m_a \lesssim 8 {\rm GeV}$, decay into tau pairs dominates; 
for $8 {\rm GeV} \lesssim m_a \lesssim 100 {\rm GeV}$, decay into bottom pairs dominates; 
and for $m_a\gtrsim 100 {\rm GeV}$, decay into gluon or gravitino pairs dominates. 
Note that for large $f_a$, the decaying into the gravitino pairs dominates other channels for the heavy R-axion mass 
since the suppression factor $f_a/M_{\rm pl}$ is not so small that overwhelms the loop factors. 
From this figure, We can see that the lifetime is shorter than $10^{13}$s for $m_a \simg 3\GeV$.



\begin{figure}[h]
\begin{center}
\includegraphics[width=.5\textwidth]{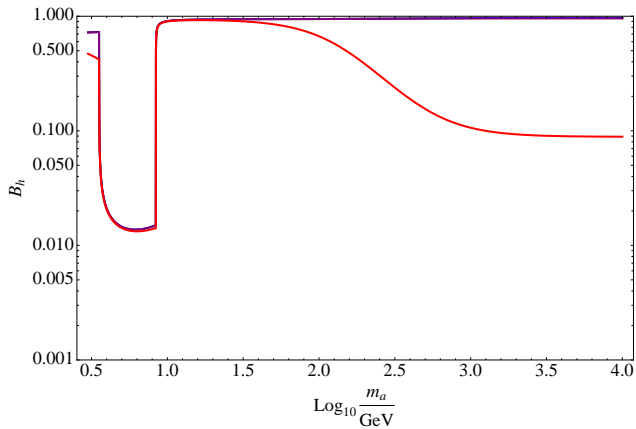}
\caption{Hadronic branching ratio of R-axion. 
Purple and red lines correspond to $f_a=10^{13}\GeV,$ and $10^{16}\GeV$. Here we use $\tan\beta=30$. If we take $f_a< 10^{13}\GeV$, we obtain the value of ${\rm B}_h$ which is the same as one for $f_a= 10^{13}\GeV$.}\label{hadron}
\end{center}
\end{figure}

In the case that the energy injected quarks are high enough, right after the axion decay, quarks and gluons immediately turn into hadronic jets\footnote{
If the axion mass is around GeV, axion dominantly decay into non- or semi-relativistic pions. 
In this case the detail of the decay process matters and the analysis is subtle and complicated, and hence
we do not consider such cases.}.
Hence, the process does not depend on the first particles created by axion decay. 
For the BBN constraint, 
then, only the branching ratio decaying into hadronic particles determines the constraint. 
We present the hadronic branching ratio of the R-axion ${\rm B}_h$ for $\tan \beta = 30$ in Fig.\ref{hadron}, 
where ${\rm B}_h$ is the sum of the branching ratio of the R-axion to colored particles. 
${\rm B}_h$ becomes small if we take larger $f_a$ since the branching ratio of the R-axion decay 
into gravitinos becomes sizable. 
It is found that ${\rm B}_h$ becomes constant for large $m_a\gtrsim 10^3{\rm GeV}$, 
$\Gamma(a \rightarrow 2g)/(\Gamma(a \rightarrow 2g)+\Gamma(a\rightarrow 2 \psi))\sim {\rm min.}\{1,  (g_3/4\pi)^4 M_{\rm pl}/f_a\}$.
Numerically ${\rm B}_h$ is of $\CO(10^{-1})$ for $f_a\simeq 10^{16}\GeV$.
For $10 {\rm GeV} \lesssim m_a \lesssim 10^2 {\rm GeV}$, decay channel into bottom pairs dominates the 
total decaying ratio, we have ${\rm B}_h\sim 1$ regardless of $f_a$. 
For $3{\rm GeV} \lesssim m_a\lesssim 10 \GeV$, ${\rm B}_h$ becomes small once more 
because the decay channel to taus dominates the total decay ratio. 
In this range ${\rm B}_h$ is of $\CO(10^{-2})$. Summarizing the above, ${\rm B}_h$ is bigger than $10^{-2}$ for $m_a \simg 3\GeV, f_a \siml 10^{16}\GeV$. 
\red{Thus, the constraint with $B_h=1$ gives the stringent constraint whereas that with $B_h=10^{-3}$ gives the conservative one, as we will see in the next section. } 

\section{Cosmological constraints on R-axion abundance\label{sec4}}

\subsection{R-axion production\label{sec3}}
Let us consider the R-axion cosmology. First we study the R-axion production. 
The R-axion production depends on cosmological scenarios, 
and we here simply suppose that the $U(1)_R$ breaking occurs after inflation. 
In this case, cosmic R-string forms at the time of the phase transition \cite{Eto:2012ij}. 
The cosmic R-strings enter the scaling regime quickly, and the cosmic string loops that are produced continuously 
emit R-axions, which is the first source of R-axions \cite{Davis:1986xc}.
Gradually the explicit $U(1)_R$ breaking effect can no longer be neglected, 
and string network turns into string-wall system, which is unstable and decay to R-axions immediately \cite{Sikivie:1982qv}.
This is the second source of R-axions.
At the same time, coherent oscillation of R-axions also starts, 
which is the third source of R-axions \cite{vacuum misalignment}.
In addition to these R-axion production from the dynamics, there are thermal production \cite{thermal production}
and production from R-saxion decay of R-axions. 
Note that it can be shown that production from R-saxion decay is
negligible (see Appendix of Ref.~\cite{Hamada:2012fr}). 
Here we summarize the R-axion production from the dynamics (the coherent oscillation, the decay of cosmic string, decay of the string-wall system) and thermal bath referring the result of Ref.~\cite{Hamada:2012fr}. 
The R-axion abundance produced from  dynamics is given by
\begin{align}
\frac{\rho_{a,{\rm dyn}}}{s}
& \simeq \left\{
\begin{array}{ll}
\red{9.4 \times 10^{-\blue{7}} {\rm GeV}  \left(\dfrac{m_a}{10{\rm GeV}}\right)^{1/2} \left(\dfrac{f_a}{10^{10}{\rm GeV}}\right)^2} &\text{for}\quad H_{\rm osc}<H_R\\
 1.7 \times 10^{-10}{\rm GeV}  \left(\dfrac{f_a}{10^{10}{\rm GeV}}\right)^2 \left(\dfrac{T_R}{10^6{\rm GeV}}\right)  & \text{for} \quad H_{\rm osc}>H_R
\end{array}\right.
\label{dynamics} ,
\end{align}
where $T_R$ is reheating temperature. $H_R$ and $H_{\rm osc}$ are Hubble parameters evaluated at the time of reheating and of beginning of the R-axion oscillation, respectively. 
Note that the result changes only by numerical factors, but parameter dependence does not change 
 if we consider the scenario without cosmic string formation but only with the coherent oscillation of R-axion \cite{Hamada:2012fr}.
The abundance of R-axions produced thermally is given by 
\al{
\frac{\rho_{a,\text{th}}}{s}&\simeq\left\{
\begin{array}{ll}
\red{2.6\times 10^{-2} \text{GeV}\left(\dfrac{m_a}{10\text{GeV}}\right)} & \text{for}\quad T_R>T_D\\ 
\red{2.0\times 10^{-2}\text{GeV} g_3^6 C_3^2  \left(\dfrac{m_a}{10\text{GeV}}\right) \left(\dfrac{10^{10}\text{GeV}}{f_a}\right)^2\left(\dfrac{T_R}{10^6\text{GeV}}\right)} & \text{for} \quad T_R<T_D
\end{array}\right. 
,
}
where $T_D=10^6{\rm GeV} g_3^{-6} C_3^{-2} \left(f_a/10^{10}{\rm GeV}\right)^2$ is the decoupling temperature. 
This value is fixed when 
R-axions become non relativistic. 
Note that for $T_R>T_D$, R-axions are once thermalized and the R-axion abundance becomes 
independent of the reheating temperature.

\subsection{Constraints for parameter space}
The standard BBN scenario can explain the light elements in the present Universe elegantly.
However, if massive exotic particle decays occur during or after
BBN epoch, light elements would be broken by the decay products, 
which would abandon the successful BBN~\add{\cite{Reno:1987qw,Kawasaki:2004qu}}.
In particular, if the decay includes hadronic decay with hadronic jets, 
a lot of $^4$He's are destroyed and $^3$He, D, and T are produced from $^4$He dissociation, 
which  gives much stringent constraint.
Hence the amount of hadronic decay product must be small enough, and in turn, R-axion abundance is constrained. 
The abundance of R-axions is constrained with respect to their decay rate and hadronic branching ratio.
For ${\rm B}_h=10^{-3}$, this constraint~\cite{Kawasaki:2004qu} is given by
\al{
\label{bh-3}
\frac{\rho_a}{s}\lesssim \left\{
\begin{array}{ll}
10^{-17/2} (\tau_a/1{\rm s})^{-5/2}{\rm GeV}  & \text{for} \quad 10^{-1}{\rm s}<\tau_a<1 {\rm s}\\ 
10^{-17/2}{\rm GeV}  & \text{for} \quad 1 {\rm s}<\tau_a<10^2{\rm s}\\ 
10^{-6}(\tau_a/1{\rm s})^{-5/4}{\rm GeV}  & \text{for} \quad 10^2 {\rm s}<\tau_a<10^4 {\rm s} \\ 
10^{-11}{\rm GeV}  & \text{for} \quad 10^4 {\rm s}<\tau_a<10^6 {\rm s} \\ 
10^{-2}(\tau_a/1{\rm s})^{-3/2}{\rm GeV}  & \text{for} \quad 10^6 {\rm s}<\tau_a<10^8 {\rm s} \\ 
10^{-14}{\rm GeV}  & \text{for} \quad 10^{8}{\rm s}<\tau_a<10^{10}{\rm s} \\ 
\end{array}\right.
.
}
%
\red{The constraint for $B_h=1$~\cite{Kawasaki:2004qu} is also given by
\al{
\label{bh1}
\frac{\rho_a}{s}\lesssim\left\{
\begin{array}{ll}
10^{-16}(\tau_a/1{\rm s})^{-5}{\rm GeV} &\text{for}\quad 10^{-2}{\rm s}<\tau_a<10^{-1}{\rm s} \\ 
10^{-23/2} (\tau_a/1{\rm s})^{-1/2}{\rm GeV}  & \text{for} \quad 10^{-1}{\rm s}<\tau_a<10 {\rm s} \\ 
10^{-12}{\rm GeV}  & \text{for} \quad 10 {\rm s}<\tau_a<10^2 {\rm s} \\ 
10^{-10}(\tau_a/1{\rm s})^{-1}{\rm GeV}  & \text{for} \quad 10^2 {\rm s}<\tau_a<10^4 {\rm s} \\ 
10^{-14}{\rm GeV}  & \text{for} \quad 10^4{\rm s}<\tau_a<10^5 {\rm s} \\
10^{-33/2}(\tau_a/1{\rm s})^{1/2}{\rm GeV}  & \text{for} \quad 10^5 {\rm s}<\tau_a<10^7 {\rm s} \\ 
10^{-6}(\tau_a/1{\rm s})^{-1}{\rm GeV}  & \text{for} \quad 10^7 {\rm s}<\tau_a<10^8 {\rm s} \\ 
10^{-14}{\rm GeV}  & \text{for} \quad 10^{8}{\rm s}<\tau_a<10^{10}{\rm s}\\ 
\end{array}\right.
.
}
}
\blue{ 
Since
we have seen $B_h\gtrsim 10^{-2}$ from Fig.~\ref{hadron}, we obtain
the conservative bound if we use \eqref{bh-3}. The larger ${\rm B}_h$
is, the more severe constraint becomes. 
Then, \eqref{bh1} for ${\rm B}_h=1$ gives more severe bound  than \eqref{bh-3}.
Note that $B_h=1$ is the good approximation for $m_a\gtrsim10\GeV$ and $f_a\lesssim 10^{13}\GeV$.  Later, we will show the results of both cases.}


There is another cosmological constraint on the R-axion abundance. 
Since R-axions can decay into the stable gravitinos, their abundance from R-axion decay may overwhelm the 
present abundance of dark matter (DM). 
Since the gravitino abundance should not exceed that of DM, we have the following constraint as
\begin{equation}
{\rm Br}_{\psi}\frac{2 m_{3/2}}{m_a}\frac{\rho_a}{s}< 4.7 \times 10^{-10} {\rm GeV} \left(\frac{\Omega_m h^2}{0.13} \right), \label{DMconst}
\end{equation}
where $\rm{Br}_{\psi}$ is branching ratio of gravitino. 
Note that the present gravitino abundance has a suppression factor ${\rm Br}_\psi (2 m_{3/2}/m_a)$, 
since a R-axion decays into two relativistic gravitinos with the total energy $m_a$ and gradually becomes nonrelativistic.
Since this constraint comes from the present Universe, it
exists regardless of the lifetime of R-axion if $\tau_a<\tau_0$.

We should also note the constraint from the gravitino produced thermally. 
The thermally produced gravitino abundance is given by \cite{Endo:2007sz}
\red{
\begin{equation}
\frac{\rho_{3/2}^{({\rm th})}}{s} \simeq 6.3 \times 10^{-10} \GeV \left(\frac{m_{\tilde g}}{10 \TeV}\right)^2 \left(\frac{m_{3/2}}{10 \GeV}\right)^{-1}\left(\frac{T_R}{10^6 \GeV}\right), 
\end{equation}
}
which must be smaller than, again,  $4.7 \times 10^{-10} {\rm GeV} (\Omega_m h^2/0.13)$, 
\red{where $m_{\tilde g}$ is the gluino mass.}
Therefore, we need relatively large gravitino mass to avoid the overclosure problem, depending on the gravitino mass. 
In other words, relatively large $f_a$ and $m_a$ are required, see Eq.~\eqref{gravmass}.

\begin{figure}[ht]
\begin{center}
\includegraphics[width=.4\textwidth]{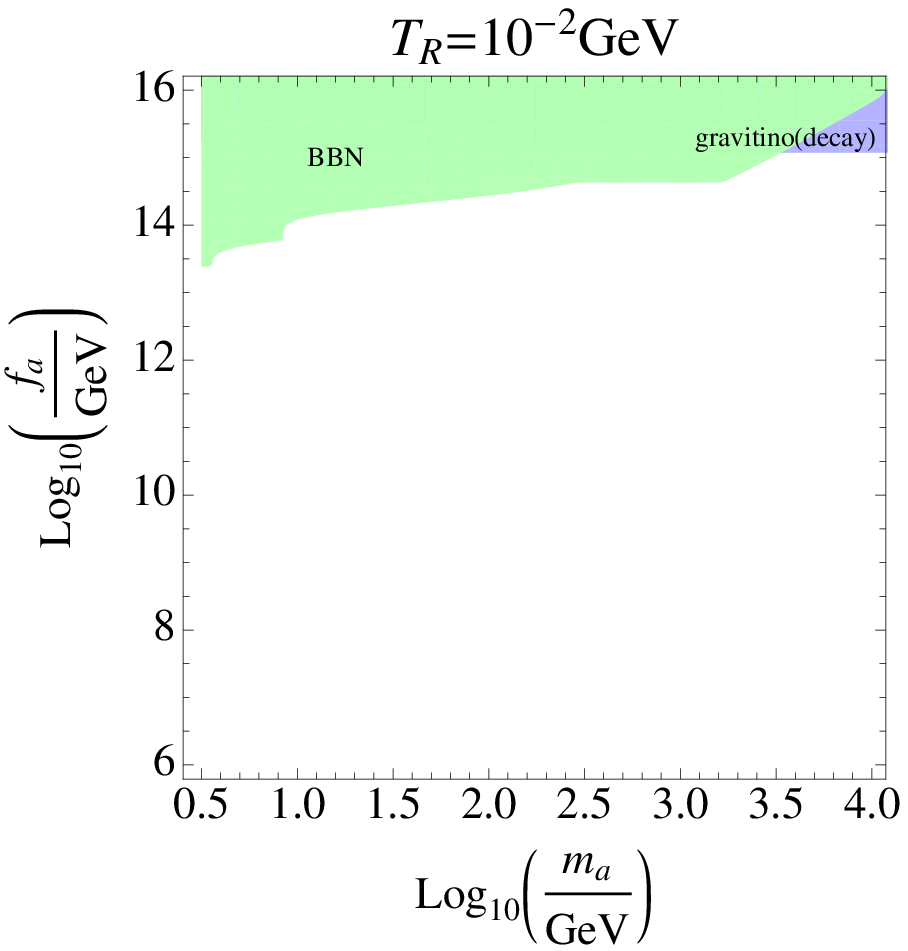}
\hfill
\includegraphics[width=.4\textwidth]{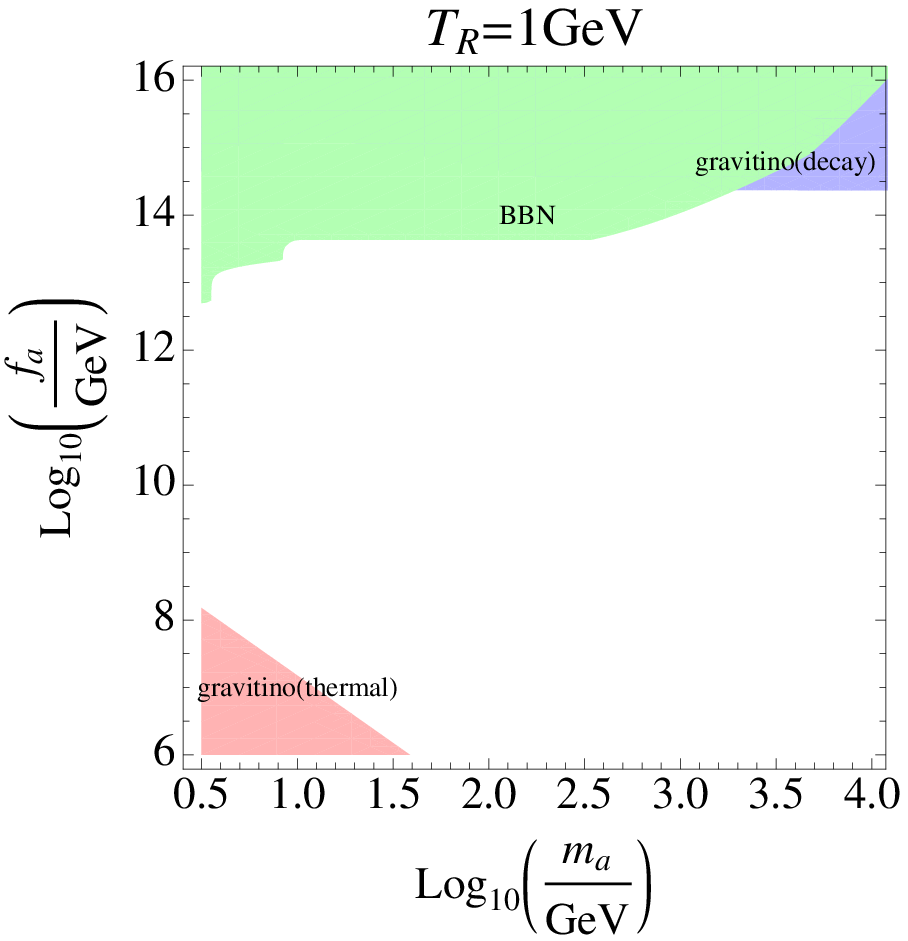}
\hfill\mbox{}
\hfill
\includegraphics[width=.4\textwidth]{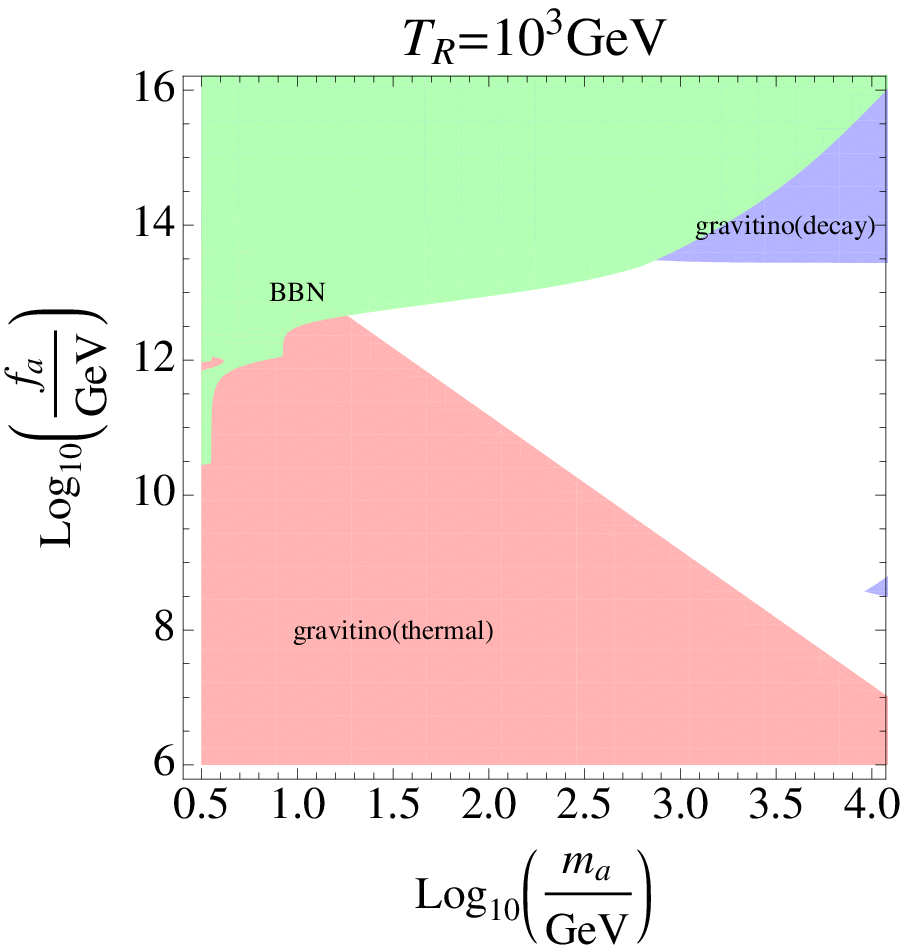}
\hfill
\includegraphics[width=.4\textwidth]{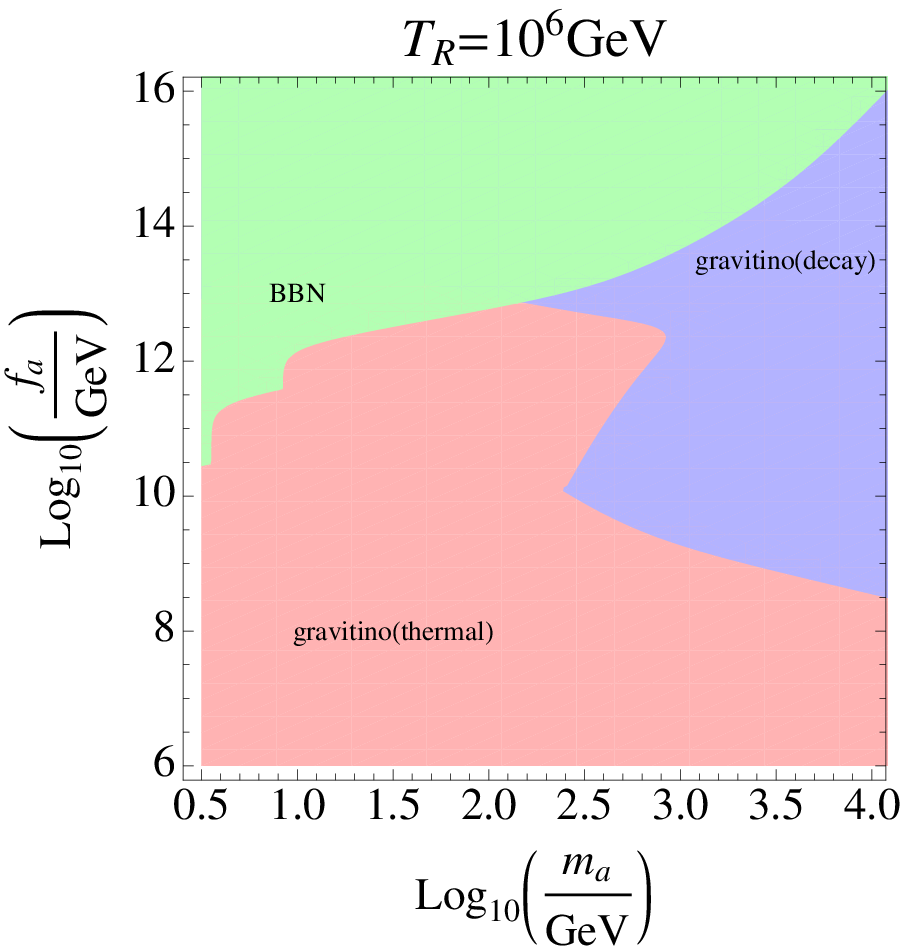}
\hfill\mbox{}
\caption{Cosmological constraints on the model parameters, $m_a$ and $f_a$ with $T_R=10^{-2}\GeV, 1\GeV, 10^3\GeV$ and $10^6\GeV$. \red{Here we use the BBN constraint \eqref{bh-3} for $B_h=10^{-3}$}. Colored region is excluded by BBN or gravitino overproduction. \red{The abundance of the gravitino produced from thermal plasma(R-axion decay)  exceeds that of DM in the blue(red) region. Here we have set $m_{\tilde g}=10 {\rm TeV}$. }}\label{constraint}
\end{center}
\end{figure}
\begin{figure}[ht]
\begin{center}
\includegraphics[width=.4\textwidth]{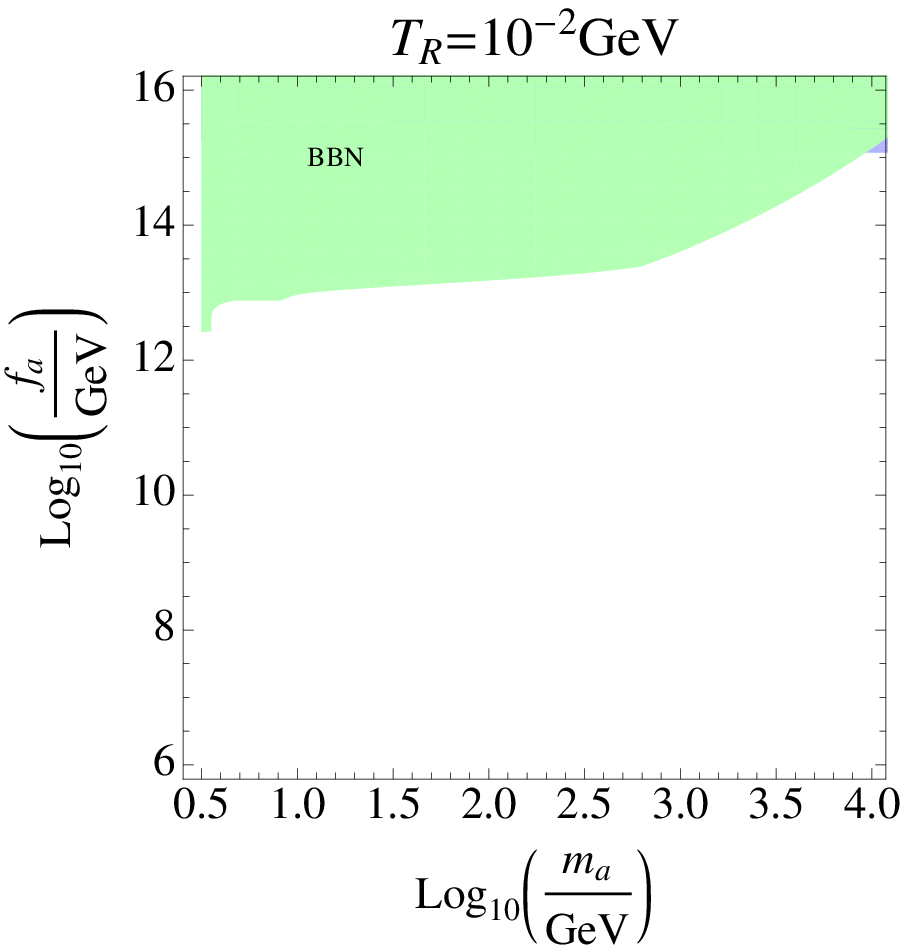}
\hfill
\includegraphics[width=.4\textwidth]{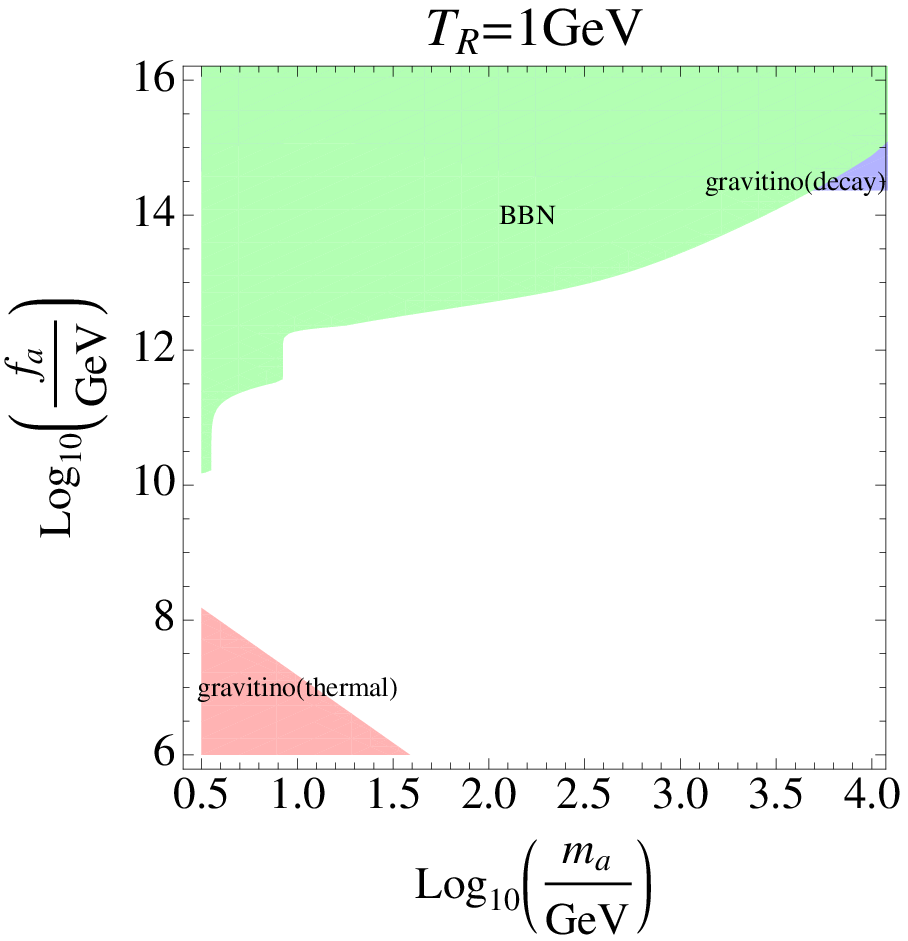}
\hfill\mbox{}
\hfill
\includegraphics[width=.4\textwidth]{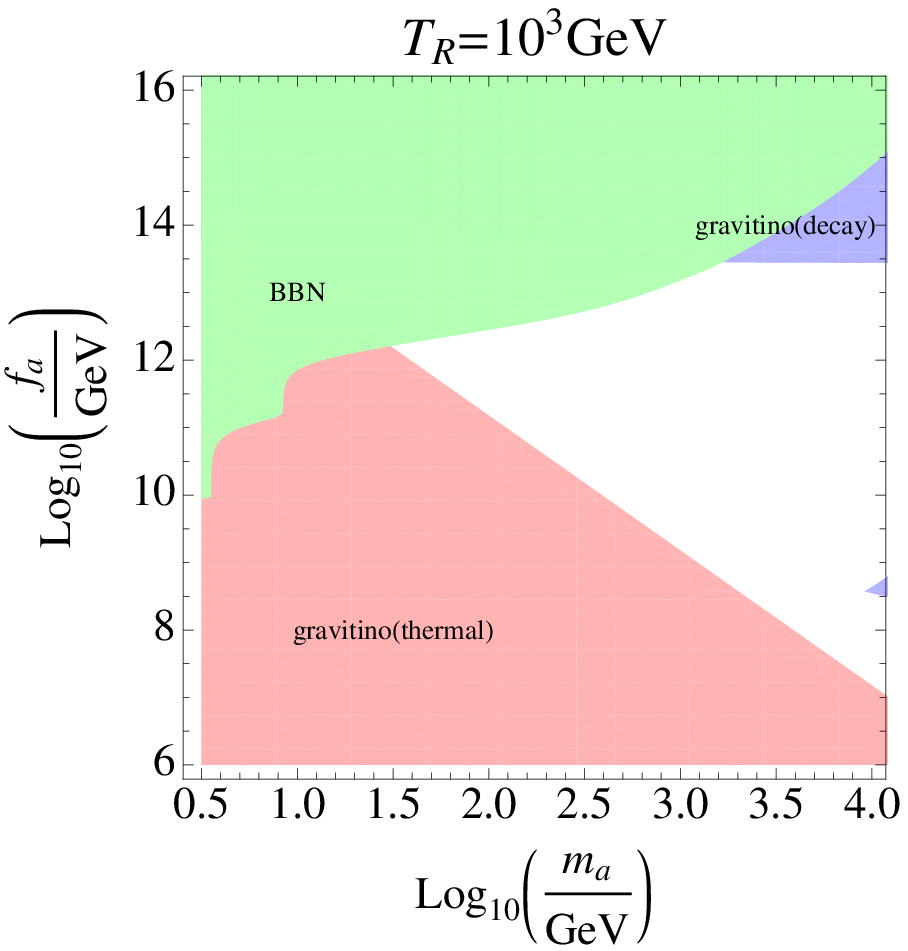}
\hfill
\includegraphics[width=.4\textwidth]{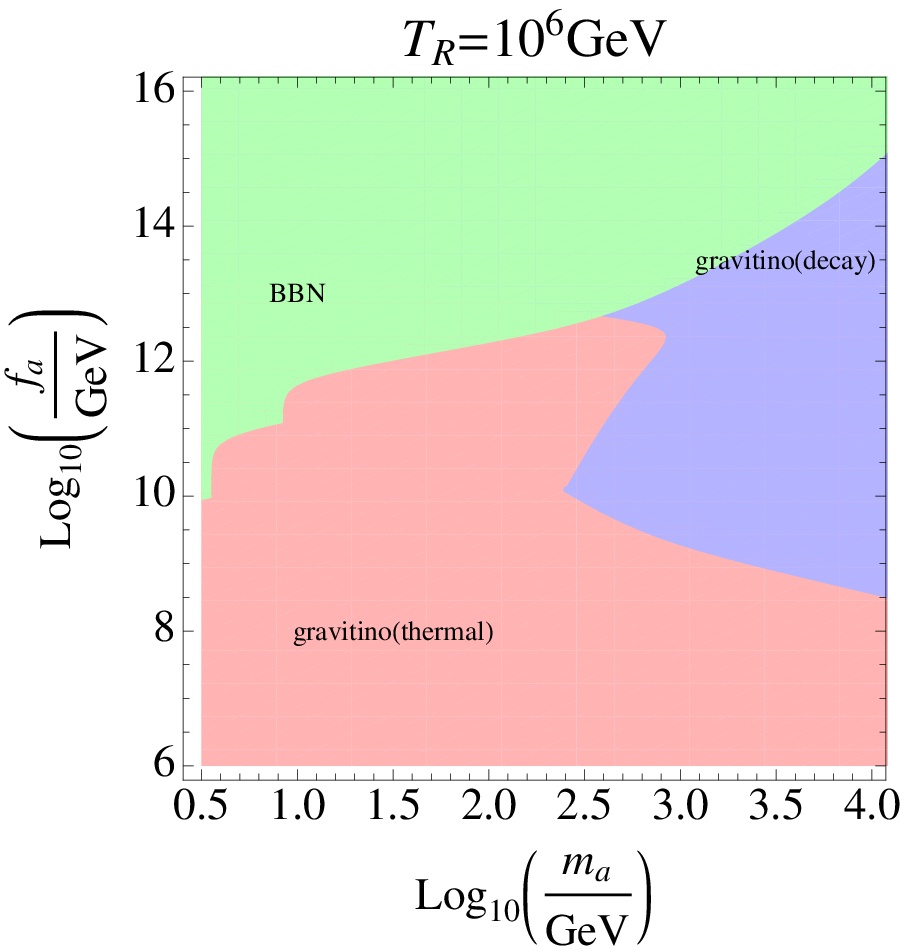}
\hfill\mbox{}
\caption{\red{Cosmological constraints on the model parameters, $m_a$ and $f_a$ with $T_R=10^{-2}\GeV, 1\GeV, 10^3\GeV$ and $10^6\GeV$. Here we use the BBN constraint \eqref{bh1} for $B_h=1$. Colored region is excluded by BBN or gravitino overproduction. The abundance of the gravitino produced from thermal plasma(R-axion decay)  exceeds that of DM in the blue(red) region. Here we have set $m_{\tilde g}=10 {\rm TeV}$. }}
\label{constraint2}
\end{center}
\end{figure}
In Fig.~\ref{constraint} and Fig.~\ref{constraint2}, we present the cosmological constraints on model parameters  discussed above, 
in terms of $m_a$ and $f_a$ with various cases of the reheating temperature.
\red{Here we have set $m_{\tilde g}=10 {\rm TeV}$. }
If the reheating temperature is high, $T_R\simeq 10^6$ GeV, thermal production of R-axion is so large that 
the gravitino abundance from the R-axion decay exceeds that of DM for $m_a\gtrsim 10^3$ GeV and $f_a\gtrsim 10^{9}$ GeV. 
Since the gravitinos produced thermally overclose the Universe for light gravitinos, which means small $m_a$ and $f_a$, 
the allowed parameter space is very small. 
Allowed parameter region enlarges for lower reheating temperature. 
In particular, the gravitino overclosure problem is almost absent for $T_R<1$ GeV. 
In this case, model parameters are constrained only by the BBN. 
We can see that R-symmetry breaking scale $f_a$ is constrained from above regardless of the reheating temperature, 
$f_a\siml10^{12-14}\GeV$. Note that 
we find that the constraint would not change so much if we use the precise value of $B_h$ (between $10^{-3}$ to 1)
by comparing the constraints of $B_h=1 $ and $10^{-3}$. 

\add{We comment on the generality of the constraints shown in Fig.~\ref{constraint} and Fig.~\ref{constraint2}. 
R-axions couple with SM gauge fields through anomaly couplings and couple with the SM fermions through the mixing between the R-axion and the Higgs bosons which comes from B-term in Higgs potential.
These couplings are general in spontaneous R-symmetry breaking models up to numerical factor.
On the other hand, it would be possible to change the gravitino mass relation Eq.~\eqref{gravmass} and the R-axion gravitino coupling Eq.~\eqref{gravitino_decay} if we consider more complicated superpotential. 
In order to avoid the constraint from gravitinos, we need a model to change Eq.~\eqref{gravmass} or Eq.~\eqref{gravitino_decay}.
}

It would be interesting to further explore heavier R-axion such as $m_a\simg 10$TeV. In this case, various decay modes into superparticles are open. Thus, arguments become highly model dependent. We will study some examples elsewhere.

\section{Summary and discussions  \label{sec6}}

In this paper, we considered the spontaneous R-symmetry breaking model and investigated the 
cosmological constraints of heavy R-axion which can decay to hadrons. 
This work complements our previous one \cite{Hamada:2012fr}. 
We estimated the abundance of the R-axion produced by decay of R-string and domain wall, vacuum misalignment and thermal plasma. 
Such abundance is constrained by BBN and the gravitino overproduction. 
We showed cosmological constraints on model parameters, R-axion mass and R-symmetry breaking scale. 
As a result, we found that $U(1)_R$ breaking scale is constrained 
as $f_a<10^{12-14}$ GeV for low reheating temperature regardless of the value of R-axion mass.
 For high reheating temperature,  BBN, gravitinos from R-axion decay, and gravitinos from 
 thermal plasma  constrain different parameter regions and the allowed parameter space is very small. 
 In conclusion, even in the heavy R-axion regime, 
it has poor compatibility with relatively high reheating temperature $T_R \gtrsim 10^6$ GeV.
\add{
The constraints we showed in this paper can apply to many spontaneous R-symmetry breaking models and are important for phenomenological model building.}

Finially it would be useful to re-interpret out results shown above as a constraint for messenger scale by using a simple gauge mediation model, taking the following simple messenger sector,
\al{
W_{\text{mess}}=\lambda' X \Phi \bar{\Phi}+M_{\Phi}\bar{\Phi}\Phi, 
}
where $\Phi$ and $\bar \Phi$ represent messenger fields. 
In this set up, the stop mass $m_0$ is given by 
\al{
m_0&=\frac{\alpha_s}{4\pi}\lambda'\sqrt{\lambda}\frac{f_a^2}{M_{\Phi}}\nn
&=\frac{\alpha_s}{4\pi}\lambda'\frac{m_a\sqrt{M_{\rm pl} f_a}}{M_{\Phi}}.
}
In the second line, we used Eq.\eqref{saxion1} and Eq.\eqref{saxion2}.
From this expression we obtain the following formula,
\al{\label{Higgs condition}
f_a\simeq 0.5 \times 10^{14} \GeV \frac{1}{\lambda'^2}\left(\frac{M_{\Phi}}{10^{11}\GeV}\right)^2\left(\frac{10\GeV}{m_a}\right)^2\left(\frac{m_0}{10\TeV}\right)^2.
}
In order to realize the appropriate Higgs mass, we take $m_0\simeq 10\TeV$ \cite{Feng:2013tvd}. 
Therefore, from Eq.\eqref{Higgs condition} 
$M_{\Phi}\siml 10^{10-11} \GeV$ is required for $m_a\simeq10\GeV, \lambda'\simeq1$ 
to obtain $f_a\siml10^{12-14}\GeV$.

\section*{Acknowledgment}

The authors would like to thank M. Ibe for useful comments and
discussions. Especially, we would like to thank K. Kohri for his
lectures on the Big Bang Nucleosynthesis and for crucial comments on
hadronic decay of R-axion. The work of Y. H. was supported by a
Grant-in-Aid for  Japan Society for the Promotion of Science (JSPS)
Fellows No.25$\cdot$1107. 
\blue{T.K. is supported in part by the Grants-in-Aid for  Scientific
  Research (A) No. 22244030  and 25400252 from the Ministry of Education, Culture,Sports, Science and 
Technology of Japan. }
The work of YO is supported by Grant-in-Aid for Scientific Research from the Ministry of Education, Culture, Sports, Science and Technology, Japan (No. 25800144 and No. 25105011). 
\red{The work of KK is supported by a JSPS postdoctoral fellowship for
research abroad.}

\appendix


\end{document}